\documentclass[twocolumn,aps,superscriptaddress,pre]{revtex4}

\usepackage{amsmath,amssymb,graphicx}
\usepackage{algorithmic}
\usepackage{enumerate}
\usepackage{times}
\usepackage{color}
\usepackage{soul}
\definecolor{yblue}{rgb}{0.06, 0.3, 0.57}
\usepackage[pdftex]{hyperref}
\hypersetup{colorlinks=true,linkcolor=blue,citecolor=blue,urlcolor=blue}

\newcommand{\mean}[1]{\langle{#1}\rangle}

\begin{document}

\title{Chaos picture of spin glasses}

\author{Wenlong Wang}
\email{wenlongcmp@scu.edu.cn}
\affiliation{College of Physics, Sichuan University, Chengdu 610065, China}

\begin{abstract}
The chaos picture is a novel, coherent, and intuitive theoretical framework for spin glasses, reconciling the 40-year-old droplet vs. replica symmetry breaking debate. First, we argue the formation of a hierarchical energy landscape by pinning and dynamic filtering of random droplets. As the temperature decreases, the chaos picture describes a progressive freezing of scale-dependent droplets in real space, which mirrors a hierarchical free energy landscape fracturing process in phase space. This bifurcation process is accompanied by temperature chaos, leading to dynamic pure state coexistence without violating the droplet scaling. Crucially, the interface geometry is not critical to these mechanisms, naturally incorporating the interface crossover from planar, fractal, to space-filling as the dimensionality increases, in line with the mean-field limit. Finally, the chaos picture provides profound insights into spin-glass dynamics such as aging, rejuvenation, and memory effect, while offering a streamlined reinterpretation of the complex metastate framework. 
\end{abstract}

\maketitle

\section{Introduction}

Spin glasses are disordered and frustrated magnets with intriguing properties and diverse applications, e.g., in biology and optimization problems \cite{Young:RMP,book}. 
The mean-field Sherrington-Kirkpatrick (SK) model \cite{SK} has an unconventional low-temperature phase of many pure states organized via the replica symmetry breaking (RSB) \cite{parisi:79,parisi:80,parisi:83}. Here, a pure state refers to a self-sustained equilibrium thermodynamic state characterized by a time-averaged spin orientational pattern $\{\mean{S_i}_t\}$. Particularly, the RSB features a hierarchical structure and is nonself-averaging. 
Despite several decades of efforts, it remains a formidable open challenge whether RSB applies to the more realistic Edwards-Anderson (EA) spin glass \cite{EA} in three dimensions. The RSB picture \cite{parisi:08,mezard:87} assumes that the mean-field theory is qualitatively correct for the EA model. On the other hand, the droplet picture \cite{mcmillan:84b,bray:86,fisher:86,fisher:87,fisher:88} based on renormalization group predicts only a single pair of pure states much like a ferromagnet. They also differ on the interface geometry (fractals or space-filling) \cite{Wang:fractal}, and the existence of a spin-glass phase in a weak external magnetic field \cite{almeida:78}. There are other pictures as well \cite{book}. The applicability of RSB is of broad interest and is related to, e.g., the Gardner transition in structural glasses \cite{charbonneau:14,Mike:GT}.

There is tremendous numerical evidence that there are many pure states for the sizes available, which have been steadily growing over time. We have been focusing on this fundamental problem in recent years, as a solid answer on one individual property can put stringent constraint on possible theories. This strategy has been pretty successful, as controversial topics are gradually clarified, they \textit{all} consistently point towards many pure states. 
Here, we briefly summarize the key progresses along this research track, including the pertinent works of others. 
To our knowledge, all numerical simulations find a nontrivial central weight under periodic boundary conditions (PBC) at a typical low temperature, see, e.g., \cite{Enzo:Review,palassini:00,Yucesoy:Delta,Wang:PA,Wang:Delta}.
This is a \textit{numerical fact}. 
The free boundary condition seemed to support the two-state picture as the central weight decays for small sizes following the $1/L^\theta$ scaling, where $\theta$ is the interface free energy exponent \cite{Katzgraber:FBC}. However, we find that this is a finite-size effect from the surfaces \cite{Wang:FBC} and the central weights of the free and periodic boundary conditions run together for larger sizes, supporting many pure states. 
Second, various statistics were proposed to support the two-state picture, but most of them are not very successful; see, e.g., \cite{Yucesoy:Delta2} and the references therein for a collection of examples. 
The fraction of centrally peaked instances was such a controversial but stimulating statistic \cite{Yucesoy:Delta,billoire:13,Yucesoy:Reply}, but this was also recently clarified \cite{Wang:Delta}. This statistic has a complicated dependence on the system size and the temperature, and it is also sensitive to the fluctuation of the central weight. Importantly, suitably modified statistics strongly suggest that the many-state picture is again most likely and coherent \cite{Wang:Delta}.
A work on thermal boundary conditions (TBC) along with a sample stiffness extrapolation argued against many states \cite{Wang:TBC}. This was partially addressed \cite{Wang:EA4D,Wang:IT}, and we shall discuss this again in this work. 
Finally, a number of pioneering works focused on zero temperature, finding a single pair of ground states \cite{Young:GS,Hatano:GS,Itoi:GS}. It is frequently believed that 
two ground states at zero temperature imply two pure states at a finite temperature and conversely many pure states imply many ground states. 
Recently, we established that a single pair of ground states is a strong support for neither a two-state nor a many-state picture, and a single pair of ground states at zero temperature and many pure states at a finite temperature are fully compatible for Gaussian disorder \cite{Wang:Delta,Wang:IT}. 
These systematic and coherent results provide compelling evidence for many pure states.

There is strong evidence that domain-wall surfaces are fractals in three and four dimensions \cite{Wang:fractal,Wang:fractal2,Wang:TC,Wang:EA4D}. Herein, we assume droplet excitations and domain-wall excitations are similar, which appears to be valid. 
Interestingly, the numerical works \cite{Wang:fractal,Wang:fractal2} find a gradual interface dimension crossover as the dimensionality is changed. The interface is planar in 1D $(d_s=d-1=0)$, then it becomes a genuine fractal $(d-1<d_s<d)$ and $d_s-d+1$ gradually increases until it becomes space-filling at an upper critical dimension in 6D $(d_s=d)$. While they are the leading-order result, they appear to be quite coherent. Particularly, they are in line with the droplet description of temperature chaos \cite{mckay:82,bray:87,Hukushima:Chaos,Katzgraber:Chaos,fernandez:13,Wang:TC,Parisi:TC,Berker:clockglass,Zhai:TC}, and the highly accurate 2D estimate \cite{weigel:dw}. Nevertheless, there are arguments to conclude space-filling in 3D by introducing correction terms in the scaling analysis \cite{Wang:DC}.


The main purpose of this theoretical work is to propose an evidence-based chaos picture, a coherent conceptual framework to reconcile the 40-year-old droplet vs. RSB debate. The chaos picture essentially describes a hierarchical free energy landscape shattering process, and the simultaneous competition of the bifurcated pure state weights featuring temperature chaos as the temperature decreases. First, the formation of a hierarchical energy landscape is caused by pinning and dynamic filtering of random droplets. As the temperature decreases, the free energy landscape fracturing takes place as scale-dependent droplets become progressively frozen. We should describe the details later in both the real space and the phase space, and discuss also why the hierarchical tree structure typically carries asymmetry. Once the ``valleys within valleys'' structure is established, we show that temperature chaos naturally enables dynamic pure state coexistence \cite{Wang:EA4D,Wang:IT}. 
In the chaos picture, typical droplet excitations are expensive, in line with the droplet picture. However, because of temperature chaos, one can accidentally register pure state coexistence, e.g., when two dominant pure states are carrying the weight transfer. This coherently explains the apparent paradox of a finite central weight and a finite stiffness exponent, they are statistical artifacts and they can perfectly coexist at the disorder-averaged level. Crucially, the core mechanisms do not depend on the interface geometry. 
Therefore, the interface crossover \cite{Wang:fractal,Wang:fractal2} is naturally accommodated within the chaos picture. 
In fact, this is a quite reasonable result. The short-range 3D spin glass lives on a cubic lattice, it is not too surprising if the droplet geometry is modified. By contrast, the long-range SK spin glass has \textit{no} geometric structure. If we arrange the spins artificially, e.g., on a ring, the large-scale excitations are inevitably space-filling and will appear everywhere. To our knowledge, the chaos picture is quite generic, it perfectly fits and offers physical insights into the RSB theory of the SK model. Finally, because the chaos picture features a hierarchical structure, it naturally explains many nonequilibrium concepts, such as aging, rejuvenation, and memory effect \cite{jonason:98,Parisi:dynamics}. The chaos picture also offers a streamlined reinterpretation of the complex metastate framework \cite{newman:97,newman:92,read:14,Parisi:Metastate} by its global chaos. 

This paper is structured as follows. First we introduce the models, and observables in Sec.~\ref{setup}. Then we discuss the theoretical framework of the chaos picture and testable predictions in Sec.~\ref{results}. Finally, we present a summary of the main findings and future challenges in Sec.~\ref{so}.

\section{Theoretical setup}
\label{setup}
We study the short-range EA and the mean-field SK Ising spin glasses \cite{EA,SK}. The EA spin glass is defined by the Hamiltonian $H = - \sum_{\langle ij \rangle} J_{ij} S_i S_j$, where $S_i=\pm 1$ are Ising spins and the sum is over nearest neighbours on a simple hypercubic lattice under PBC. For a linear size $L$, there are $N=L^d$ spins. The random couplings $J_{ij}$ are chosen independently from the standard Gaussian distribution with mean zero and variance one. A set of quenched couplings $J=\{J_{ij}\}$ defines a disorder sample. The 2D model has a spin-glass phase at $T=0$ \cite{weigel:dw}. The 3D model has a finite-temperature phase transition at $T_C \approx 1$ \cite{katzgraber:06}. 
If the PBC is switched to antiperiodic boundary conditions (APBC) in one spatial direction, a relative domain wall is created. 
In the TBC \cite{Wang:TBC,Wang:EA4D}, multiple competing boundary conditions are present, e.g., the system can choose between PBC or APBC along some or all spatial directions according to their weights. For convenience, TBC2 signifies that the TBC is applied in one spatial direction. 
The SK spin glass is defined similarly but on a fully connected graph. The Hamiltonian reads $H = - 1/\sqrt{N} \sum_{i<j} J_{ij} S_i S_j$. The transition temperature is exactly known $T_C=1$. Because of the continuous bonds, there is a unique pair of ground states \cite{Young:GS,Hatano:GS,Itoi:GS,Wang:IT}. 

The spin-glass order parameter is the spin overlap distribution function $P_J(q)$ where the spin overlap $q$ is defined as:
\begin{align}
    q_{\alpha\gamma} =\frac{1}{N}\sum\limits_i\mean{S_i}_\alpha\mean{S_i}_\gamma.
\end{align}
Here, we use lowercase Greek letters to label pure states. In simulations, the spin overlap is measured as: 
\begin{align}
    q=\frac{1}{N}\sum\limits_i S_{i}^{(1)} S_{i}^{(2)}, \label{q}
\end{align}
where spin configurations ``(1)'' and ``(2)'' are chosen independently
from the Boltzmann distribution. The central weight of an individual overlap function is defined as:
\begin{align}
    I_J(q_0) = \int_{-q_0}^{q_0} P_J(q) dq.
    \label{I02}
\end{align}
Here, $q_0$ characterizes the half length of a chosen interval around $q=0$. We take the common choice $q_0=0.2$ unless otherwise specified. This quantity is of fundamental importance. First, this central region is essentially insensitive to the influence of the $q_{\mathrm{EA}}$ peaks at a typical low temperature. Second, because the $q$ values are small, it represents genuinely system-size excitations. Similarly, if a peak appears in this window, we term it as a central peak. When the subscript is dropped, we refer to the disorder-averaged quantity. 

The link overlap is defined similarly as:
\begin{equation}
q_l=\dfrac{1}{N_B}\sum\limits_{\langle ij \rangle} S_{i}^{(1)}S_{j}^{(1)}S_{i}^{(2)}S_{j}^{(2)},
\label{ql}
\end{equation}
where $N_B$ is the number of bonds. For example, $N_B=dN$ and $N(N-1)/2$ for the EA and SK models, respectively. While the spin overlap focuses on the volume of excitations, the link overlap is related to the surface of excitations. It should be noted that the link overlap does not measure the level of satisfied bonds, rather it compares the patterns of the (un)satisfied bonds. Naturally, it is frequently relevant for the estimation of interface dimensions.

The $d_s$ for the EA model is defined as $\Sigma \sim L^{d_s}$, where $\Sigma$ is a measure of the surface area of the excitation. The stiffness exponent $\theta$ is defined as $\Delta F \sim L^\theta$, where $\Delta F$ is the interface free energy. See \cite{Wang:SKD} for a generalization of $d_s$ and $\theta$ to long-range models, including the SK model. The SK large-scale excitations are space-filling and the stiffness exponent $\theta>0$, including bond-diluted models \cite{Wang:fractal,Wang:fractal2,Wang:KAS,Wang:SKD}. 
Finally, the regular thermodynamic observables such as the energy, free energy, and the entropy can be calculated almost as continuous functions of the temperature for individual samples \cite{Wang:IT}. 

\section{Chaos Picture}
\label{results}
First, we define and discuss some subtle concepts to avoid ambiguity. We also discuss the conceptual difficulties of the droplet picture on a single pair of pure states. Second, we formulate the chaos picture. We start from the formation of a hierarchical rugged energy landscape. Next, we describe the free energy landscape shattering, and the dynamic pure state coexistence by temperature chaos, as the temperature decreases. Finally, we discuss the interface dimension crossover, nonequilibrium phenomena, the metastates, and a few testable predictions of the chaos picture.

\subsection{Remarks and motivations}
We define the local equilibrium and the global equilibrium of spin glasses. A low-temperature pure state is a dynamically trapped macrostate in the phase space, and it is in local equilibrium in its own valley. Intuitively, a pure state encompasses all the microstates it can dynamically visit in principle, either experimentally or computationally with local updates, e.g., a single-spin update. 
The weight of a pure state is therefore the sum of the Boltzmann factors of its microstates.
In the spin-glass phase, a pure state features an apparently random spatial pattern and a time ordering symmetry breaking. 
In this vein, a nonequilibrium relaxation of a spin glass is a convergence process to a particular pure state in the spin-glass phase. 
A global equilibrium is more like a theoretical construct, aiming to overview the entire symmetry breaking pattern. Since each pure state has a weight, the ensemble of all pure states with their weights is defined as the global equilibrium. Indeed, if we normalize the weight of a pure state by the partition function, it becomes the probability of this pure state in the ensemble. In global equilibrium, a sample can visit the entire phase space, in spite of the symmetry breaking. 
In simulations, this can be achieved by cluster updates or extended ensemble Monte Carlo methods. 

We distinguish between the total number and the effective number of pure states in global equilibrium.
This can be crucial for certain topics to avoid ambiguity. The former is by counting, it is the total number of pure states, regardless of their weights. The latter reflects the number of pure states that actually have an $O(1)$ probability. To this end, we define the effective number of pure states as $N_{\mathrm{eff}}=\exp(-\sum_\alpha p_\alpha\ln(p_\alpha))$, where $p_\alpha$ is the probability of pure state $\alpha$ \cite{Wang:IT}. Here, we refer to the dominant pure states as active pure states. The pure states with vanishing probabilities are referred to as either permanently extinct pure states or dormant pure states. An extinct pure state after its formation has a vanishing weight for all lowering temperatures. By contrast, a dormant pure state temporarily has a vanishing weight, but it becomes an active one or even the dominant one as the temperature decreases. Importantly, many pure states do not automatically imply many active states. In fact, there is frequently a single pair of active states, and a special example is the ground states at $T=0$ \cite{Wang:IT}. It should be emphasized that the overlap function counts the weights, i.e., it essentially measures the overlap of active states in practice. For example, the overlap function at $T=0$ is simple, despite that there is a massive ergodicity breaking in the phase space.


It is widely accepted that spin glasses have a rugged (free) energy landscape. However, the meaning of this depends on the context. In the RSB picture, the energy landscape itself is highly rugged with a hierarchical structure. In the droplet picture, there are two dominant pure states, but each basin has a rough landscape. While the droplet picture posits a globally simple landscape featuring a single pair of pure states, its prediction of temperature chaos requires a radical, large-scale reorganization of the ground state under infinitesimal temperature shifts. This sensitivity raises a conceptual difficulty: if macroscopic droplet flips are continuously triggered by marginal parameter variations, the static ``pair of valleys" from the ground state attractors becomes rather blurred. Indeed, a finite-temperature pure state pattern may look radically different from ``its'' ground state attractor. In addition, if we measure the spin overlap function exactly in the temperature chaos process, we should register pure state coexistence. To our knowledge, there is no obvious mechanism to prevent this from happening \cite{Wang:EA4D,Wang:IT}. Therefore, the droplet picture itself strongly suggests pure state coexistence by its own prediction of temperature chaos. Indeed, it seems likely that $\theta>0$ does not appear to exclude pure state coexistence and imply two states \cite{Wang:EA4D,Wang:Delta,Wang:IT}. Particularly, the $\theta$ exponent is a growing function of dimensionality \cite{boettcher:05d} and even the SK model has a positive exponent \cite{Wang:KAS,Wang:SKD} which clearly has many pure states. 
Our chaos picture resolves these tensions by suitably modifying the droplet picture. In the chaos picture, the apparent paradox of a finite central weight and a finite stiffness exponent is a statistical artifact at the disorder-averaged level.

\subsection{Formation of a hierarchical energy landscape}
As mentioned earlier, the chaos picture describes essentially a hierarchical free energy landscape fracturing process, and the simultaneous competition of pure state weights featuring temperature chaos as the temperature decreases. Therefore, the formation of a hierarchical energy landscape is central to the chaos picture. 
In fact, once this is established, the emergence of many pure states is quite straightforward by temperature chaos, as we should discuss later.

Here, we provide a heuristic but highly reasonable argument by combining the increased suppression of droplets of correlation length scale $\xi(T)$ in the real space, and the probabilistic emergence of a few new attractors in the state space by pinning and dynamic filtering. Interestingly, this naturally leads to a tree structure in the symmetry breaking. While a rigorous mathematical proof of this multi-scale topology in finite dimensions remains elusive, this framework naturally bridges the gap between the scale-dependent properties of the droplet model and the hierarchical topology of RSB.

First, the chaos picture exhibits a real-space, scale-dependent freezing of droplets. 
We reiterate that a pure state is fundamentally a dynamic concept on its existence, and the free energy cost to flip a droplet is not a concern, which affects its thermodynamic weight. 
The system first breaks the global $Z_2$ symmetry at full system size scale at $T_C$. Near the critical temperature $T_C$, the divergence of the correlation length naturally implies that relevant excitations are macroscopic, scaling as the system size $O(L)$. As the temperature is lowered, it is physically expected that the characteristic length scale of these excitations should shrink, as the correlation length decreases. At very low temperatures, only very small local excitions remain active as $\xi(T)$ converges to lattice scale. We argue that this progressive reduction in the active excitation length scale acts as the driving mechanism for further real-space symmetry breaking and a hierarchical landscape in phase space. Rather than merely modifying a static ground state, the shrinking spatial boundaries force a sequence of localized breaks in phase space—consecutively nesting smaller ``valleys within valleys" as large-scale flips freeze out. 

An acute reader may find this argument also applies to the ferromagnet, why there are only two valleys there but multiple valleys may form for a spin glass? Furthermore, there are numerous random droplets of length scale $\xi(T)$ for pure states bifurcation. There are two crucial effects here. First, the disorder and frustration lead to the pinning effect, creating dynamic barriers. 
Suppose we arbitrarily flip a droplet and create a relative domain wall, if there is a surface corner spin with some strongly satisfied interior bonds alongside an equal number of weakly frustrated surface bonds (all bonds were satisfied before the flip), this spin will be highly unwilling to yield, acting as a directional pinning site that blocks the domain wall from moving backward. 
Furthermore, even if a transient fluctuation forces this corner spin to flip, it is highly likely to quickly flip back to realign with its dominant local field, particularly at low temperatures. 
For the droplet to shrink and evaporate, the domain wall must systematically overcome a massive number of such pinning sites with aligned local fields, and naturally the number of barriers is scale dependent. 
Crucially, this pinning struggle intensifies dramatically in higher dimensions, where the increased spatial connection and coordination number heavily reinforce the collective grip of the interior bonds, naturally generating stable dynamic traps of droplet excitations. 
Second, not all arbitrary droplet excitations of $\ell \sim \xi(T)$ can function as attractors in the state space. 
To function as an attractor, a configuration must exhibit structural stability against a continuous spectrum of localized thermal perturbations, ranging from single-spin flips to small-cluster rearrangements. 
This stringent criterion acts as a powerful dynamical filter, eliminating most arbitrary droplet flips. Combining them together, a small number of stable valleys \textit{may} form probabilistically, i.e., they are not guaranteed to exist deterministically in every single disorder realization at all temperatures, which is also not required. Instead, the hierarchical tree structure manifests stochastically as an emergent property of sample-dependent, temperature-dependent dynamical trapping, in line with the nonself-averaging of RSB.



\begin{figure*}[tbh]
\begin{center}
\includegraphics[width=0.872\columnwidth]{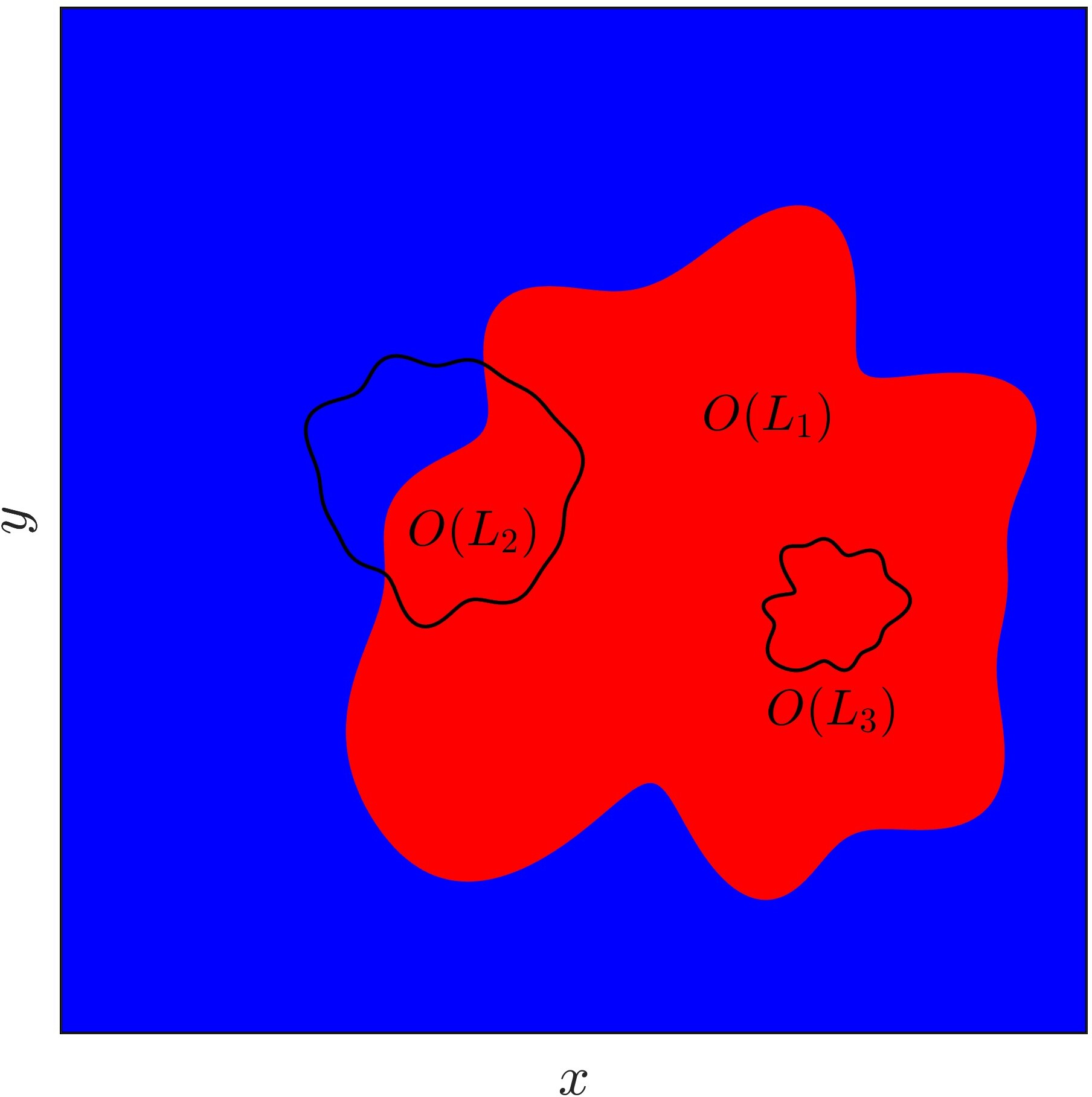}
\put (-195,200) {$(a)$} 
\includegraphics[width=\columnwidth]{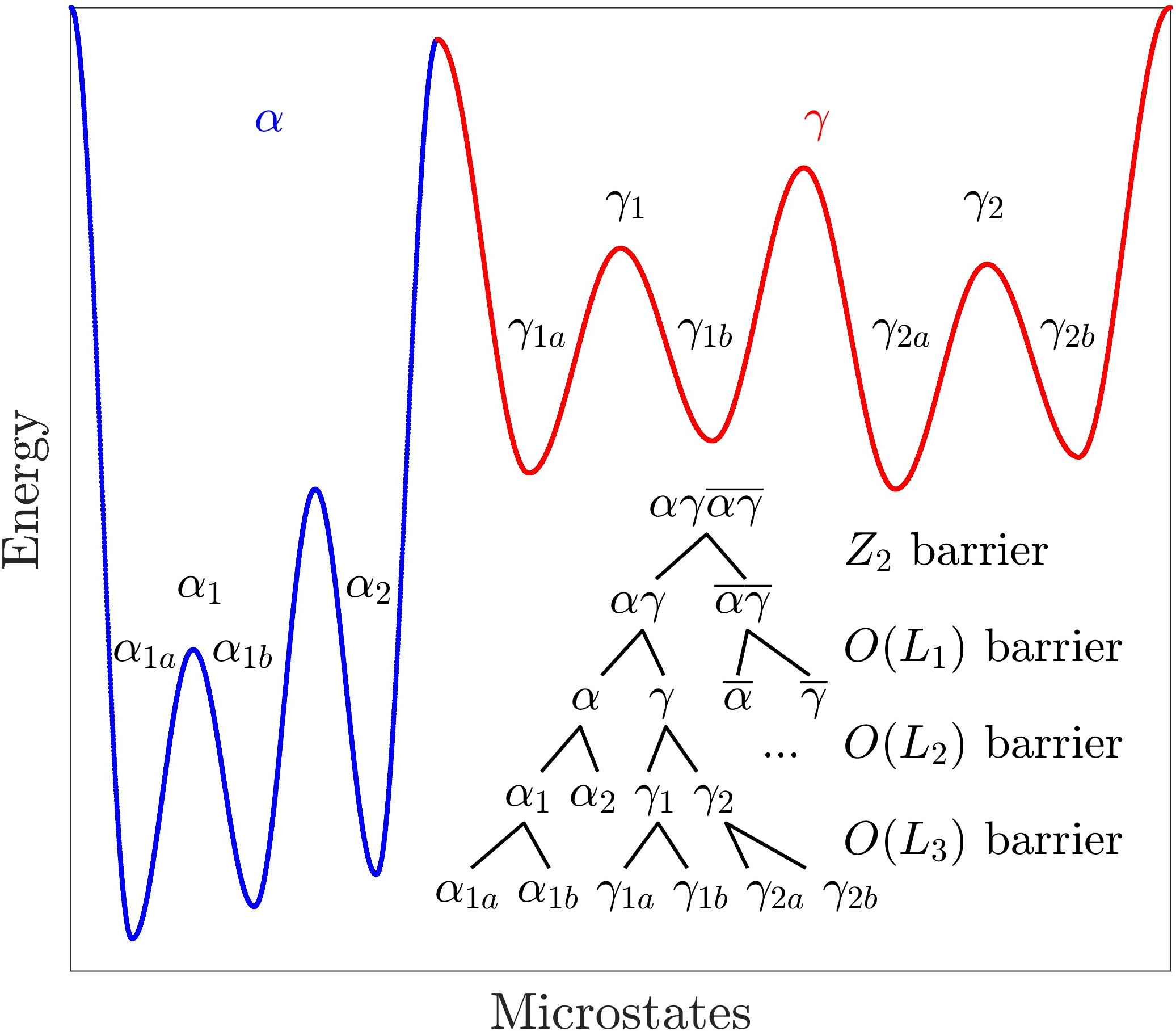}
\put (-224,200) {$(b)$} 
\caption{
A schematic diagram on the formation of the hierarchical energy landscape, illustrated in both the real space (a) and the phase space (b). 
The system first breaks the $Z_2$ symmetry, see the $Z_2$ barrier on the sides. As the temperature decreases, a large red cluster becomes frozen, leading to the the $\alpha$ (not flipped) and $\gamma$ (flipped) pure states. Here, we show the $\gamma$ state relative to the $\alpha$ state in real space for clarity. At a lower temperature, a straddling cluster further splits $\gamma$ into $\gamma_1, \gamma_2$ pure states, but this cluster is absent for $\alpha$ and the tree becomes asymmetric. However, $\alpha$ may split into $\alpha_1, \alpha_2$ by another cluster of similar size passing through both the $O(L_1)$ and $O(L_3)$ clusters (not shown in real space, which focuses on the $\gamma$ branch). Finally, the $O(L_3)$ cluster becomes frozen and it is relevant to $\alpha_1, \gamma_1, \gamma_2$, leading to a further splitting of these pure states at a yet lower temperature. The phase space diagram also illustrates temperature chaos. At a higher temperature, the high-energy, high-entropy $\gamma$ is the dominant state. As the temperature decreases, it gradually transfers the weight to the low-energy, low-entropy $\alpha$ pure state. In this process, pure state coexistence is achieved, despite that typical $\alpha, \gamma$ free energy difference is macroscopic by droplet scaling. See the text for further details.
}
\label{Tree}
\end{center}
\end{figure*}

Crucially, the hierarchical structure in phase space (valleys within valleys) does not translate to simple spatial nesting (droplets within droplets) in real space. It is helpful to illustrate this with an example, and a schematic bifurcation process is depicted in both the real space (left) and the phase space (right) in Fig.~\ref{Tree}. The system first breaks the (outer) $Z_2$ barrier, followed by the $O(L_1)$ barrier, which separates the $\alpha$ valley and the $\gamma$ valley. Here, the droplet is highlighted for the $\gamma$ pure state relative to the $\alpha$ pure state. When the $O(L_1)$ droplet is flipped and becomes pinned and frozen, a smaller straddling droplet $O(L_2)$ may freeze at a lower temperature for the $\gamma$ state, yielding the $\gamma_1, \gamma_2$ pure states. It is perfectly possible that it pins well, in fact it may even share pinning sites with the $O(L_1)$ droplet, as they may share common surface segments at the lattice scale. 
Before the bifurcation, it can flip back and forth, it is sufficiently large to trigger a subsequent bifurcation and also sufficiently small to maintain the essential feature of the $\gamma$ pure state and therefore stay in the $\gamma$ branch. At a yet lower temperature, the $O(L_3)$ droplet flip is frozen, and both $\gamma_1, \gamma_2$ undergo a further bifurcation as this small droplet does not interfere with the two larger ones.
The ultrametric tree architecture is a dynamical property of the configurations in the state space topology, seamlessly coexisting with overlapping, complex fractal interfaces in real space.



The bifurcation tree structure may become asymmetric by the broken local symmetry caused by consecutive droplet flips. Consider the two pure states $\alpha, \gamma$ in Fig.~\ref{Tree} again, while the $O(L_2)$ droplet is relevant for the $\gamma$ pure state symmetry breaking at a lower temperature, it is not necessarily relevant to $\alpha$. 
Indeed, this local symmetry is broken for a straddling droplet. For example, this broken local symmetry may manifest dynamically through the mechanism of shared boundary pinning. As mentioned above, the $O(L_2)$ droplet may not intersect the $O(L_1)$ droplet arbitrarily at the lattice scale; rather, it may opportunistically share localized segments of the $O(L_1)$ interface. In this way, they may share pinning sites that are entirely absent in the $\alpha$ state. In this scenario, it is possible that $\gamma$ splits into $\gamma_1, \gamma_2$, while the pure state $\alpha$ stays quite rigid in this process. 
This configuration-dependent pinning completely decouples the branching histories of the two valleys. 
In our diagram, the $\alpha$ state undergoes a similar bifurcation but by a different droplet of similar size, which is not shown for clarity. This droplet intersects both the $O(L_1)$ and $O(L_3)$ droplets, such that the subsequent bifurcation by $O(L_3)$ is relevant to the $\gamma_1, \gamma_2, \alpha_1$ states, but not the $\alpha_2$ valley. This asymmetry is clearly present for the SK model, therefore, it seems likely that the asymmetric branching is a fundamental signature of the spin-glass phase, whether viewed from the abstract mean-field theory or the real-space fractal interfaces. 



\subsection{Temperature chaos leads to dynamic pure state coexistence}

Here, we show that pure states can coexistent by temperature chaos in the framework of the hierarchical energy landscape, despite that typical free energy difference scales as $L^\theta$. Temperature chaos describes weights transfer between pure states. The competition of pure states is vividly illustrated by the (partial) TBC framework \cite{Wang:TC,Wang:BC,Wang:EA4D}, in which different boundary conditions compete and their weights oscillate as the temperature decreases. While this illustrates the competition of groups of pure states, the basic idea is nevertheless very similar.

Temperature chaos is very natural for the asymmetric, hierarchical energy landscape. Consider the $\alpha, \gamma$ pure states in Fig.~\ref{Tree} before they split further, the high-energy, high-entropy $\gamma$ state may dominate at a higher temperature and as the temperature decreases, it transfers the weight to the low-energy, low-entropy $\alpha$ state at a lower temperature. Naturally, we expect that there is a temperature interval for this power transfer process. In this interval, the probability of $\gamma$ decreases from $1$ to $0$, and the probability of $\alpha$ increases from $0$ to $1$. While their ``typical" free energy difference both before and after the chaos process is enormous, the free energy difference is actually $O(1)$ rather than macroscopic in this pertinent interval. In fact, there is a temperature at which their probabilities are exactly equal and remarkably $\Delta F=0$ for a massive droplet flip. It should be highlighted that this arises purely from a delicate balance of their free energies as the temperature is tuned, no symmetry transformation is required to link the two pure states. If one happens to measure the $P_J(q)$ in this temperature interval, one may see a central peak and register an ``RSB-like" sample. Clearly, if the working temperature barely missed this interval, one may register it as a ``droplet-like'' sample. The two pure states have no particularly desire to have or maintain the $O(1)$ interface free energy. This power transfer is relevant throughout the entire low-temperature phase, in parallel with the bifurcation process. If $\alpha_2$ had a lower energy, a similar power transfer might happen from $\alpha_1$ to $\alpha_2$. Similarly, if there were a very narrow but deep pure state $\delta$, there could be a power transfer from $\alpha_{1a}$ to $\delta$, and a central peak might emerge even at very low temperatures. Indeed, it is observed that the average central weight is very small but finite for both the 3D EA and SK spin glasses at a very low temperature $T=0.01$ \cite{Wang:IT}. At a typical low temperature, most samples are ``droplet-like" but some samples are ``RSB-like", the chaos picture provides a natural explanation as some samples are undergoing temperature chaos while the others are not at the observing temperature. While we illustrated the idea with two pure states, the essential feature remains if multiple pure states coexist. Finally, we mention that the system is not physically crossing the barriers in temperature chaos, the equilibrium chaos is a consequence of the global equilibrium.

The chaos picture is a highly dynamic picture, and pure state coexistence may appear and then disappear as the temperature decreases. One prediction of the chaos picture is that individual central weight oscillates in the low-temperature phase, and this was numerically confirmed in \cite{Wang:IT}. 
Our picture also explains the peak splitting phenomena of the temperature evolution of $P_J(q)$, see the movie of \cite{Wang:IT}. This is a direct signature of the free energy landscape fracturing. While two active pure states register a (relatively broad) peak in the chaos process, it is possible that one or less likely both undergo further symmetry breaking, yielding a splitting of the peak. For example, if $\gamma$ splits into $\gamma_1, \gamma_2$ first, its overlap with $\alpha$ will split into two peaks. The splitting is also possible for a symmetry breaking of a single pure state for its self-overlap. For example, if $\alpha_1$ splits into $\alpha_{1a}, \alpha_{1b}$, the original $q_{\mathrm{EA}}$ peak would split into three peaks. As the temperature decreases, the thermal fluctuation weakens and (active) pure states tend to become more localized in the phase space, and this explains the peak-sharpening process as the temperature declines for a finite system \cite{Wang:Delta}. The state space of TBC is enlarged by the macroscopic degrees of freedom, this should generate a even more rugged energy landscape rather than simplifying it. This enhances temperature chaos, and active pure states of different boundary conditions tend to be more different, leading to a larger central weight \cite{Wang:TBC}.


A chaotic process between two active pure states is analogous to a first-order phase transition. It is tempting to argue that the transition happens instantaneously in the thermodynamic limit. However, a regular first-order transition has only one transition. While a single process does take a narrower temperature interval $O(1/L^\zeta)$, where $\zeta=d_s/2-\theta$ is the chaos exponent, the number of such transitions grows exactly as $L^\zeta$, completely compensating the narrowing effect \cite{Wang:EA4D,Wang:IT}. Therefore, the chance to register chaotic processes of active pure states is essentially size independent, and this is the fundamental root of the approximately constant central weight \cite{Enzo:Review,palassini:00,Yucesoy:Delta,Wang:PA,Wang:Delta}. The $O(1)$ interface free energies of the chaotic process have a rather minor effect on the droplet scaling, which is dominated by costly excitations. The chaos picture coherently explains the coexistence of a finite central weight and a finite stiffness exponent at the disorder-averaged level \cite{Wang:IT}.



\subsection{Droplet geometry, nonequilibrium dynamics, metastates, and more testable predictions}

The core mechanisms of the preceding discussion barely depend on the precise nature of the droplet geometry. Therefore, the interface dimension crossover can be seamlessly incorporated into the chaos picture \cite{Wang:fractal,Wang:fractal2}. The $d_s-d+1$ gradually increases from $0$ in 1D to $1$ at the upper critical dimension estimated as 6D, i.e., the interface is planar in 1D, space-filling in 6D, and a fractal in between. A similar crossover is found in the framework of the long-range, power-law decay Kotliar-Anderson-Stein spin glass. Particularly, the interface is space-filling in the mean-field regime \cite{Wang:fractal,Wang:fractal2,Wang:SKD}. 
As mentioned earlier, they are the leading-order result, but they appear to be quite coherent, e.g., they are in good agreement with the droplet description of temperature chaos \cite{mckay:82,bray:87,Hukushima:Chaos,Katzgraber:Chaos,fernandez:13,Wang:TC,Parisi:TC,Berker:clockglass,Zhai:TC}, and the highly accurate 2D estimate \cite{weigel:dw}. 
We emphasize again that the interface geometry is essentially a separate property from the number of pure states in the chaos picture; cf. the metastate pictures below. 
While the TNT picture similarly predicts many pure states and fractal interfaces in three dimensions, its formulation remains largely heuristic \cite{houdayer:00b,krzakala:00,palassini:00}. The chaos picture provides a microscopic foundation within the regular droplet scaling, maintaining a single stiffness exponent across both the droplet and domain-wall excitations.

In fact, the fractal interface interpolation is a quite reasonable scenario. 
This geometric modification is relevant as the short-range EA spin glass is defined on a lattice. By contrast, the long-range SK spin glass is a fully connected model and it essentially has no geometric structure. As mentioned earlier, it is rather intuitive that its excitations are space-filling from this perspective. It is quite comforting that fractal surface converges to space-filling at a finite upper critical dimension.

The chaos picture naturally interprets central concepts of nonequilibrium dynamics with the temperature-dependent tree structure. The tree terminals typically expand or melt as the temperature decreases or increases, respectively. 
In this framework, spin glass aging depends on the initial configuration, which determines the starting tree branch. The aging process is a dynamic random walk in the phase space and a stochastic selection of the tree branches, and the final relaxation to a particular tree terminal of the working temperature. This is a strictly running down process along the tree structure, generating the spin-glass arrow of time. If the system reaches local equilibrium in a pure state at $T_1<T_C$, and the temperature is suddenly lowered to $T_2$, the tree structure expands, forcing the system to age anew, and this is the rejuvenation phenomenon. This process may continue to a yet lower temperature $T_3$. Conversely, if we raise the temperature back to $T_1$, the tree terminals melt, allowing the system to trace back up the tree structure and return to its starting valley, which is the memory effect. This retrieval is very natural, as the system has never truly left the initial bulky valley throughout the entire process. This interpretation is not unique to the chaos picture, it is a fundamental consequence of the temperature-dependent hierarchical tree structure.

The chaos picture clarifies a historical ambiguity regarding window observables in spin glass simulations. It is well-known that an embedded window of size $W$ essentially registers an identical disorder-averaged $[P_J(q)]_J$ distribution to an isolated system of the same size. We argue that treating the embedded window as an independent energy landscape is fundamentally misleading. Geometrically, if a large system under PBC is partitioned into an $n \times n$ grid of windows, the global overlap is simply the spatial mean of the window overlaps $(q_{\mathrm{sys}} = \langle q_{\mathrm{win}} \rangle$). Under the disorder average, translation invariance dictates that no window is privileged, meaning each local window acts strictly as an unbiased estimator of the global bulk state ($[q_{\mathrm{win}}]_J = [q_{\mathrm{sys}}]_J$). The window remains virtually passive; the intermediate states it registers do not stem from internal local chaos, but are the real-space projections of macroscopic, system-wide droplets slicing through its boundaries. Chaos is inherently a nonlocal, collective phenomenon of the global system.


From the perspective of global chaos, the necessity of the inspiring window description is no longer obvious. Indeed, the metastate pictures frequently employ a distinct and highly complicated lexicon, as the setup essentially relies on a changing window embedded within a varying system of vastly larger scale. The chaotic pairs picture is a two-state picture with respect to the window, yet it morphs into a many-state picture when considering chaotic size dependence \cite{newman:97,newman:92}. The nonstandard RSB picture is inherently a many-state picture, featuring also chaotic size dependence \cite{newman:97,read:14,Parisi:Metastate}. Crucially, both frameworks require space-filling excitations, otherwise, changes in the surrounding bulk would fail to affect the embedded window in the thermodynamic limit. By contrast, our global perspective dictates that as the system size varies, the entire global energy landscape is constantly altered, despite that the random bonds within and immediately around the window remain strictly fixed. The window acts merely as a passive, local observatory tracking a series of effectively distinct systems. Indeed, it is well known that even a highly localized boundary perturbation from PBC to APBC in a single spatial direction is sufficiently powerful to reshape the entire energy landscape \cite{Parisi:TC,Wang:TBC2}. Therefore, variations in system size effectively act as a proxy for a disorder average over the window via the mechanism of bond chaos. While these pioneering metastate pictures are definitely interesting, our framework strongly suggests that they can be entirely reinterpreted within a regular, finite-dimensional scaling setting that naturally accommodates fractal interfaces.




Finally, we propose several testable predictions for future numerical and experimental exploration. 
\begin{enumerate}
    \item In the chaos picture, many pure states and costly excitations coexist at the disorder-averaged level. Typical excitations are expensive following the droplet scaling $\Delta F \sim L^\theta$, but $O(1)$ excitations exist by temperature chaos. The chaos picture therefore modifies the droplet scaling slightly. The domain-wall free energy $\Delta F$ distribution has $O(1)$ contributions ($L$-independent) on top of the $L^\theta$ distribution. This should have observable consequences on the cumulative distribution function of $\Delta F$. While a preliminary test was made \cite{Wang:IT}, a more systematic study is necessary as this distribution function converges slowly. 
    \item Dynamic pure state coexistence should imply dynamic domain wall coexistence, as each boundary condition encompasses many pure states. Therefore, a fraction of samples under TBC should remain not stiff, i.e., not a single boundary condition dominates even in the thermodynamic limit. This possibility was actually conjectured in \cite{Wang:TBC}. In this vein, the observed central weight of the TBC should be larger than its PBC counterpart even in the thermodynamic limit, see \cite{Wang:TBC}. A simpler setup is to study the TBC2 in 3D or 4D with two competing boundary conditions, they should exhibit essentially similar behaviours.
    \item A pinned large droplet and then a pinned smaller straddling droplet at a lower temperature should be observable in real space, especially in dynamics. Chaos picture may provide a coherent, real-space explanation for the characteristic quakes or intermittent energy deviations observed in spin-glass dynamics. For example, they may occur when the system is on the edge of a symmetry breaking, with respect to a pertinent droplet.
    \item Since fractal interfaces are perfectly compatible with many pure states, it is now highly relevant to systematically reexamine the $d_s$ works in the literature, particularly in 3D and 4D, as what we have done on the number of pure states. The goal is to understand these works in a similarly intuitive and consistent manner. We predict that the leading-order result via a regular fit qualitatively reflects the asymptotic behaviour.
\end{enumerate}
A large-scaling Monte Carlo simulation of the three-dimensional spin glass under TBC2 is currently in progress to test the chaos picture \cite{Wang:TBC2}.

\section{Summary and outlook}
\label{so}
We presented the chaos picture, a novel, coherent, and intuitive theoretical framework for spin glasses that reconciles the forty-year-old debate between the droplet model and RSB. We argue how a ``valleys within valleys" hierarchical energy landscape is sculpted via the pinning and dynamic filtering of random droplets. As the temperature decreases, the chaos picture describes a progressive freezing of scale-dependent droplets in real space, and a corresponding hierarchical fracturing of the free-energy landscape in phase space, and temperature chaos, which leads to dynamic pure state coexistence. The chaos picture explains that many pure states and costly droplet excitations can coexist at the disorder-averaged level. Crucially, the core mechanisms are independent of interface geometry, naturally accommodating an interface crossover from planar to fractal, and ultimately to space-filling surfaces as the dimensionality increases. Finally, the chaos picture provides unified, profound insights into nonequilibrium spin-glass dynamics, while offering a streamlined reinterpretation of the complex metastate framework. 

The proposed testable predictions present compelling avenues for future exploration. If the chaos picture is correct, a natural next step is to extend this framework to include an external magnetic field and reexamine the existence of the de Almeida–Thouless (AT) line \cite{almeida:78}. Similarly, a comprehensive review synthesizing the vast and often conflicting literature on the AT line is highly valuable \cite{Mike:AT,AJ:AT}. Research efforts along some of these directions are currently in progress and will be reported in forthcoming publications.




\textit{Acknowledgments--} 
We thank M. A. Moore for helpful discussions.
We gratefully acknowledge supports from the National Science Foundation of China under Grant No. 12004268, and the Fundamental Research Funds for the Central Universities, China. 

\bibliography{Refs}

\end{document}